\documentstyle[preprint,aps]{revtex}

\tightenlines

\newcommand{\lsim}[1]{
\setlength{\unitlength}{12pt}
\begin{picture}(1.4,1.)
\put(.7,-0.3){\makebox(0.0,1.)[t]{$<$}}
\put(.7,-0.3){\makebox(0.0,1.)[b]{$\sim$}}
\end{picture}#1}
\newcommand{\gsim}[2]{
\setlength{\unitlength}{12pt}
\begin{picture}(1.4,1.)
\put(.7,-0.3){\makebox(0.0,1.)[t]{$>$}}
\put(.7,-0.3){\makebox(0.0,1.)[b]{$\sim$}}
\end{picture}#2}
\begin{document}
\draft

\title{Pulsar velocities due to a violation of the equivalence principle by
neutrinos}

\author{R.Horvat \\
   ``Rudjer Bo\v skovi\' c'' Institute, P.O.Box 1016, 10001 Zagreb,
Croatia}

\maketitle

\begin{abstract}

We consider the pulsar velocity problem and relate it to some unconventional
neutrino oscillation mechanisms based on a violation of the equivalence
principle by neutrinos. We show that the observed pulsar velocities may be
explained by violations at the level from $10^{-9}$ to $10^{-10}$ in the case 
of a non-universal tensor neutrino-gravity coupling, whereas there is no 
solution in the case of a non-universal scalar neutrino-gravity coupling. 
Neutrinos may remain massless and the requisite magnetic field strength  is
similar to that in the conventional mass oscillation mechanism.
\end{abstract}

\newpage

   Recently it has been proposed by Kusenko and Segr\`{e} that high peculiar
velocities of pulsars, of the order of a few hundreds of kilometers per
second \cite{1}, may be explained by a mechanism \cite{2} based on resonant 
neutrino conversions. According to these authors the origin of the observed
velocities of pulsars lies in the angular dependence of the neutrino
resonance surface, caused by the strong background magnetic field present
during a SN explosion. On the one hand, the mechanism is very elegant as the
anisotropy in the neutrino emission may be achieved even for matter density
which is spherically symmetric, whereas, on the other hand, it is less elegant
as it points to the physics beyond the Standard Model by providing some
insight into neutrino masses. 

   Considering typical pulsar masses and the fact that more than $99\%$ of
the binding energy is emitted in form of neutrinos, it is easy to estimate
that only a $1\%$ asymmetry in the distribution of the escaping neutrinos
could generate the observed kick of the pulsars. This asymmetry could be
generated by the mechanism of Kusenko and Segr\`{e} \cite{2}, if  inside
the protoneutron star there is a magnetic field of strength $B\sim 2\times
10^{15}-10^{16}$ G \cite{3}. The basic ingredients in this mechanism are the
dispersion relation for massless neutrinos biased by the strong magnetic
field \cite{4} and the value of the neutrino mass squared difference, $\Delta
m^{2} \sim 10^{4} \;\mbox{\rm eV}^{2}$, no matter whether the resonant 
conversion is between $\nu_e$ and $\nu_{\mu}$ or $\nu_e$ and $\nu_{\tau}$. 
Later on, Kusenko and Segr\`{e} found that ordinary-sterile resonant neutrino
oscillations may also play a role in the understanding of the pulsar kicks
\cite{5}. Now, a value of the neutrino mass squared difference which is
required is  somewhat
larger, $\Delta m^{2} \sim 1 \;\mbox{\rm keV}^{2}$. But in any of the above 
mass oscillation mechanism (MOM), the value of $\Delta m^{2}$ is in conflict with
cosmological bounds on neutrino masses, unless $\nu_{\mu}(\nu_{\tau})$ is
unstable. Also, these values of $\Delta m^{2}$ are not preferred by neutrino
oscillation solution neither of the solar neutrino problem nor the
atmospheric neutrino anomaly.

   At the same time Akhmedov, Lanza and Sciama \cite{6} proposed a hybrid
mechanism which combines the resonant spin-flavor precession of neutrinos
possessing a transition magnetic moment with the mass mechanism. This
mechanism requires $B > 2 \times 10^{16}$ G as long as the Majorana
transition magnetic moment exceeds 
$\mu_{\nu} \gsim \;\mbox{\rm 10}^{-15} \mu_B$. Recently,
Grasso, Nunokawa and Valle \cite{7} have proposed a viable scheme for 
generating pulsar space velocities by considering resonant oscillations of 
massless neutrinos in a particle physics model based on a small violation of
universality in the weak interaction sector. Let us finally mention that
some possible explanations of the pulsar velocity problem (PVP) still rely 
on the Standard Model of particle physics. The main ingredient in these models  
is the parity violation in weak interactions \cite{8} or the asymmetric
collapse \cite{9}. 

   In the present note we try to relate the high birth velocities of
pulsars with a different kind of neutrino oscillations based on the
assumption that neutrinos violate the equivalence principle. The basic
mechanism, in which the neutrino mixing is generated by a violation of the
equivalence principle induced by a breakdown of universality in the
gravitational coupling strength between the conventional spin-2 particles
and the neutrinos, was proposed first in \cite{10} and discussed afterwards 
in several papers over the past ten years (see e.g. \cite{11} and the 
references therein). The alternative mechanism, in which the neutrino mixing is
generated due to a breakdown of universality in the coupling strength
between the massless spin-0 particles (the string dilaton field of
\cite{12}) and the neutrinos, has been put forward recently \cite{13}. In the 
first example (hereafter, the tensorial VEP mechanism), neutrino oscillations 
will take place even for massless neutrinos whereas the second mechanism 
(hereafter, the scalar VEP mechanism) requires that neutrinos are massive but 
not necessarily non degenerate. Let us now follow the main steps from 
\cite{2} to test the relevance of the first mechanism mentioned above for the 
pulsar kicks.

   When the magnetic field is absent, a resonance for massless neutrinos
induced by the tensorial VEP mechanism in a two-flavor mixing scenario can 
occur at a distance $r_0$ from the center of a protoneutron star if 
\begin{equation}
2E\phi (r_0) \delta f = \sqrt{2} \,G_F N_e (r_0)\;\;\; ,
\label{form1}
\end{equation}
for $\cos{2 \theta_G} \simeq 1$, where $\theta_G$ is the mixing angle which
relates the weak and the gravitational basis. Here $G_F$ is Fermi's constant,
$N_e$ represents the electron number density, $\phi$ is the Newtonian
gravitational potential and the difference $\delta f \equiv f_2-f_1$
characterizes a VEP. Therefore, for $\delta f > 0$ the neutrinos can go
through a resonance as $\phi$ has been considered to be positive as in
\cite{10}. To estimate $\delta f$ relevant for the PVP,
we need to know the function $\phi (r)$. For that purpose we solve the
equation
\begin{equation}
\Delta \phi (r)=-4 \pi G_N \rho (r)\;\;\;
\label{form2}
\end{equation}
using the density profile ansatz:
\begin{equation}
\rho (r)= \rho_c \;(R_c/r)^m\;\;\;,
\label{form3}
\end{equation}
with  $\rho_c = 8 \times \;\mbox{\rm 10}^{14}\;g/cm^{3}$, $R_c =
10 \; km$ and $m=5-7$. In  Eq.(2) $\Delta$ is the Laplacian and $G_N$ is 
Newton's constant. These parameters were found to satisfactorily describe
the basic properties of SN1987A \cite{14}. The solutions are:
\begin{equation}
\phi^{(m=5)}(r)= \frac{G_N M}{r}\;\left( \frac{5}{2}-\frac{1}{2}
\frac{R_{c}^{2}}{r^2} \right) \;\;\; ;
\label{form4}
\end{equation}
\begin{equation}
\phi^{(m=7)}(r)= \frac{G_N M}{r}\;\left( \frac{7}{4}-\frac{3}{20}
\frac{R_{c}^{4}}{r^4} \right) \;\;\; .
\label{form5}
\end{equation}

   If $r_0$ is to lie between the $\nu_e$ neutrinosphere and the
$\nu_{\mu}(\nu_{\tau})$ neutrinosphere, as was required in the original
model of Kusenko and Segr\`{e}, then we find that 
\begin{equation}
\phi^{(m=5)}(r_0) \simeq 0.17-0.24 \;\;\; ; \;\;\; \phi^{(m=7)}(r_0) 
\simeq 0.12-0.17
\;\;\; ,
\label{form6}
\end{equation}
and by using $<E_{\nu_{e}}> \approx \;\mbox{\rm 10-12 MeV}$ and 
$Y_e \approx 0.1$ 
consequently $\delta f \simeq 10^{-10}-10^{-9}$. The confrontation of 
this favored range for the PVP to the constraints  from 
accelerator experiments will be discussed bellow.

   In order to obtain a maximal $\nu_e \rightarrow \nu_{\mu}(\nu_{\tau})$
conversion the resonant transition must be sufficiently adiabatic. The 
 condition for adiabatic transition in the tensorial VEP mechanism 
can be found to be \cite{15} 
\begin{equation}
\frac{\sqrt{2} \; G_F N_{e}^{0}\; {\tan^{2}{2 \theta_G}}}
{\left | {h}_{N_{e}}^{-1} - {h}_{\phi}^{-1} \right |} \gg 1 \;\;\;,
\label{form7}
\end{equation}
where the subscript 0 denotes that this quantity should be evaluated at the 
resonance and the scale heights are given by 
$h_{N_{e}}=|d\;\mbox{\rm ln}N_{e}/dr|_{0}^{-1}$ 
and $h_{\phi_{e}}=|d\;\mbox{\rm ln} \phi/dr|_{0}^{-1}$. Note that since the
oscillation length increases with $E$ in the tensorial VEP mechanism, the
resonant condition (7) is violated for low energy neutrinos, in
contrast to the MOM where it is violated for high
energy neutrinos. We find, roughly, that the oscillations can be treated as
adiabatic if $\sin{\theta_G} \gsim \mbox {\rm 10^{-4}}$.

   The distortion of the effective $\nu_{\mu}(\nu_{\tau})$ neutrinosphere is
best described by the dimensionless parameter \cite{3}
\begin{equation}
\alpha = \left( \frac{e}{2} \right) \left( \frac{N_e}{{\pi}^{4}} \right)
^{\frac{1}{3}} B \left( \frac{1}{N_e} \right) \left/ \;
 \left| {h}_{N_{e}}^{-1} - {h}_{\phi}^{-1} \right| \right. \;\;\;.
\label{form8}                                         
\end{equation}
Note that the adiabatic condition, Eq.(7), and the parameter $\alpha$,
Eq.(8), differ from the corresponding quantities in the MOM
 by the occurrence of the term ${h}_{\phi}^{-1}$ in the denominator
as well as $\theta_G$ instead of a mixing angle which relates the neutrino
weak interaction eigenstates and the mass eigenstates. This has a
consequence that the required value of the magnetic field strength in the
MOM is now lowered approximately by a factor 
${h_{N_{e}}}/{h_{\phi}}$ (note that $h_{N_{e}} < h_{\phi}$ ). By ignoring
the fact that $Y_e$ in fact decreases with $r$ between the two neutrinospheres
, one finds that the maximal reduction in the magnetic field strength can be 
at the very most  $20\%$. 

   We briefly mention the possibility that actually the coupling between
gravity and neutrinos might be replaced by a dependence on $\nabla
\phi $ instead of $\phi $. Such type of coupling may arise in string theory
\cite{12}. In this case the replacement of the type \cite{15}
\begin{equation}
2E \phi (r) \delta f \rightarrow E R_f |\nabla \phi |
\label{form9}
\end{equation}
must be done in the above expressions, where $R_f$ with dimension of length
now characterizes a VEP. We find that
$R_f \simeq 10^{-3} - 10^{-2} cm$ may be relevant for the PVP. Also, the
maximal reduction in the magnetic field strength can be about $40\%$ in this
case. Let us finally mention an intriguing  possibility that such kind of
coupling may solely be responsible for the anisotropic neutrino emission,
without invoking strong magnetic fields. For a distinct gravitational source
outside the protoneutron star, the coupling of neutrinos to
$\nabla \phi $ involves a dependence on the angle between the
neutrino's momentum and the position vector of the gravitational source. We
find, however, that such a coupling is hopelessly small to account for a
$1\%$ anisotropy in the distribution of the escaping neutrinos. 

   Under the hypothesis that neutrino oscillation data can solely be
explained by the tensorial VEP mechanism, the various neutrino oscillation
laboratory experimental data were analyzed in Ref. \cite{16}, including the
recent LSND observations \cite{17} for $\bar{\nu}_{\mu} \rightarrow 
\bar{\nu}_e$
oscillations, to constrain the relevant parameter space.
The authors found \cite{16} that the allowed region of BNL E776 experiment
combined with the LSND result may reveal a very small allowed parameter
region at $\sin^{2}{2 \theta_G} \approx 10^{-3}$ and $|\phi \delta f| \approx
10^{-16}$. Here $\phi$ denotes the value of the Newtonian potential at
Earth, usually taken to be dominated by the mass distribution on supercluster
scales (the great attractor). It may be prudent, however, to take $\phi$ to
be dominated by a mass distribution on scales less than supercluster as the
precise value of $\phi$ from supercluster scales is still uncertain. 
Surprisingly, if we take the value of $\phi$ which is dominated by the 
galaxy or Virgo cluster ($\phi \simeq 6 \times 10^{-7}$ and $1 \times 10^{-6}$,
respectively, see e.g. table 1 in \cite{11}), one finds that $\delta f$ fits
exactly in the range relevant for the PVP as found
before. Presumably, the small allowed window in the relevant parameter space
will be closed \cite{16} on the inclusion of  other experimental data 
\footnote{The constraints from reactor experiments, including the recent 
CHOOZ results \cite {18}, are unimportant here since these experiments are 
done with neutrinos having energies in the MeV range.}, meaning that the 
gravitational coupling of $\nu_e$ and $\nu_{\mu}(\nu_{\tau})$ are the same. 
Then, by naturalness  argument the gravitational coupling of 
$\nu_{\tau}(\nu_{\mu})$ should be equal to these, making thereby the 
resolution of the PVP impossible within the tensorial VEP 
oscillation mechanism. However, the scenario may be saved by admitting a 
nontrivial neutrino mass matrix. In this case the parameter space is 
amplified with $\Delta m^{2}$ and $\sin ^{2}{\theta}$, and the authors of 
Ref.\cite{16} argued that there would be some minimal $\Delta m^{2}$ which is 
consistent with all laboratory constraints. Evidently, we need 
$\Delta m^{2} \ll 10^{4} eV^{2}$ for $\theta \sim \theta_G $ or 
$\Delta m^{2} \gsim~10^{4} eV^{2}$ is still allowed if $\theta \gg \theta_G $. 
The latter possibility may occur as $\theta $ and $\theta_G $ are completely 
unrelated and the left hand side of (1) contains now an extra term, 
$(\Delta m^{2}/2E)\cos {2\theta}$.  
   
   Let us finally consider the unconventional way to generate neutrino
oscillations via the scalar VEP mechanism \cite{13},
 in connection to the PVP. This mechanism
requires that neutrinos are massive, but even  if the neutrinos are
completely degenerate the mechanism will still produce oscillations. The
main ingredient here is the effective mass $m^*$ in a constant gravitational
field, induced by the spin-0 exchange contribution to the static
gravitational energy. The oscillation length is obtained from that
appropriate for the MOM simply by the replacement $\Delta m^{2} \rightarrow
\Delta m^{*2}$. Therefore, the mechanism could be relevant  for the pulsar
kicks if $\Delta m^{*2} \sim 10^{4} eV$. Concerning ourself to the case
where the mass and the gravitational eigenstates are identical, this
explicitly turns out to be (for completely degenerate neutrinos)
\begin{equation}                                          
\Delta m^{*2}=-2m^{2}_{\nu} \; \alpha_{ps}  \phi(r_0)  \delta \alpha 
\sim 10^{4} eV^{2}\;\;\;,
\label{form10}
\end{equation}
where $m_{\nu}$ is the degenerate electron-neutrino mass, $\alpha_{ps}$ is
the strength with which the dilaton field couples to the matter inside a
protoneutron star and $\delta \alpha$ now characterizes a VEP in the
neutrino sector. Taking $m_{\nu} \lsim 5 eV$, $\phi \simeq 0.2$ and a limit
on $\alpha_{ps}$ to coincide to that coming from solar-system gravity
experiments \cite{19}, $\alpha^{2}_{ps} < 10^{-3}$, one finds that $|\delta
\alpha | \gsim~3 \times 10^{4}$. Since this is unacceptable large, one may
therefore conclude that the scalar VEP neutrino oscillation mechanism is
unlikely to be responsible for the large birth velocities of the neutron
stars.

   In conclusion, we have investigated some unorthodox neutrino oscillation
mechanisms in the light of
the pulsar velocity problem. We have shown that neutrino resonant
conversions due to a non-universal tensor neutrino-gravity coupling may be
responsible for generating pulsar velocities from anisotropic neutrino
emission in a strong static magnetic field, if the equivalence principle is
violated by neutrinos at the $10^{-10}-10^{-9}$ level. For the case where
neutrinos couple to the gradient of the gravitational potential, the
dimensional parameter that characterizes a VEP should be
$10^{-3}-10^{-2}cm$. Although, in principle, the neutrinos could remain
massless in this scheme, constraints from future accelerator experiments may
require a nontrivial neutrino mass matrix.

The author acknowledges the support of the Croatian Ministry of Science and
Technology under the contract 1 -- 03 -- 068.


\begin{thebibliography}{160}
\bibitem{1} A. G. Lyne and D. R. Lorimer, Nature (London) 369, 127 (1994).
\bibitem{2} A. Kusenko and  G. Segr\`{e}, Phys. Rev. Lett. 77, 4872 (1996).
\bibitem{3} Y. Z. Qian, Phys. Rev. Lett. 79, 2750 (1997). 
\bibitem{4} J. C. D'Olivo, J. F. Nieves and P. B. Pal, Phys. Rev. D40, 3679 
(1989).
\bibitem{5} A. Kusenko and  G. Segr\`{e}, Phys. Lett. B396, 197 (1997).
\bibitem{6} E. Kh. Akhmedov, A. Lanza and D. W. Sciama, Phys. Rev. D56, 6117
(1997).
\bibitem{7} D. Grasso, H. Nunokawa and J. W. F. Valle, hep-ph/9803002.
\bibitem{8} C. J. Horowitz and G. Li, Phys. Rev. Lett. 80, 3694 (1998).
\bibitem{9} A. Burrows and J. Hayes, Phys. Rev. Lett. 76, 352 (1996).
\bibitem{10} M. Gasperini, Phys. Rev. D38, 2635 (1988); D39, 3606 (1989).
\bibitem{11} A. Halprin, C. N. Leung and J. Pantaleone, Phys. Rev. D53, 5365
(1996).
\bibitem{12} T. Damour and A. M. Polyakov, Nucl. Phys. B423, 532 (1994).
\bibitem{13} A. Halprin and C. N. Leung, Phys. Lett. B416, 361 (1998).  
\bibitem{14} M. S. Turner, Phys. Rev. Lett. 60, 1797 (1988).
\bibitem{15} J. N. Bahcall, P. I. Krastev and C. N. Leung, Phys. Rev. D52,
1770 (1995).
\bibitem{16} R. B. Mann and U. Sarkar, Phys. Rev. Lett. 76, 865 (1996).
\bibitem{17} C. Athanassopoulos et al (LSND collaboration), Phys. Rev. Lett.
77, 3082 (1996); Phys. Rev. C54, 2685 (1996).
\bibitem{18} M. Apollonio et al, Phys. Lett. B420, 397 (1998).
\bibitem{19} R. D. Reasenberg et al, Astrophys. J. 234, L219 (1979).
\end{thebibliography}
\end{document}